\documentclass[12pt,preprint]{aastex}

\shorttitle{Old Populations in the Mid-UV}

\newcommand{\ang}{\textrm{\AA}}
\newcommand{\wrange}[2]{\ensuremath{ {#1} < \lambda < {#2} \, {\ang}}}
\newcommand{\teff}{\ensuremath{T_{\textrm{\scriptsize eff}}}}
\newcommand{\msun}{\ensuremath{\, M_\odot}}

\newcommand{\sigobs}{\ensuremath{\sigma_{\textit{\scriptsize obs}}}}
\newcommand{\kelvin}{\ensuremath{\, {\textrm K}}}

\newcommand{\fxv}{\ensuremath{ {\bigl( F_{1500}/F_V\bigr) }} }
\newcommand{\fxvh}{\ensuremath{F_{1500}^{\textsc{\scriptsize hot}}}}
\newcommand{\fxvc}{\ensuremath{F_{1500}^{\textsc{\scriptsize cool}}}}
\newcommand{\fvvh}{\ensuremath{F_V^{\textsc{\scriptsize hot}}}}
\newcommand{\fvvc}{\ensuremath{F_V^{\textsc{\scriptsize cool}}}}
\newcommand{\fxxvh}{\ensuremath{F_{2500}^{\textsc{\scriptsize hot}}}}
\newcommand{\fxxvc}{\ensuremath{F_{2500}^{\textsc{\scriptsize cool}}}}
\newcommand{\fxxvvh}{ \ensuremath{\Bigl({F_{2500} \over F_V}\Bigr)^{\textsc{\scriptsize hot} }}}
\newcommand{\fxxvvc}{ \ensuremath{\Bigl({F_{2500} \over F_V}\Bigr)^{\textsc{\scriptsize cool} }}}
\newcommand{\fxvvh}{ \ensuremath{\Bigl({F_{1500} \over F_V}\Bigr)^{\textsc{\scriptsize hot} }}}
\newcommand{\fxvvc}{ \ensuremath{\Bigl({F_{1500} \over F_V}\Bigr)^{\textsc{\scriptsize cool} }}}
\newcommand{\fxxv}{\ensuremath{{\bigl( F_{2500}/F_V\bigr)}}}

\begin{document}

\title{Age and Abundance Discrimination in Old Stellar Populations\\
Using Mid-Ultraviolet Colors}

\author{Ben Dorman\altaffilmark{1,2}}
\email{Ben.Dorman@gsfc.nasa.gov}
\author{Robert W. O'Connell\altaffilmark{1}, Robert T. Rood\altaffilmark{1}}
\email{rwo@virginia.edu, rtr@virginia.edu}

\altaffiltext{1}{Astronomy Dept, University of Virginia, P. O. Box 3818,
Charlottesville, VA 22903-0818}
\altaffiltext{2}{Laboratory for High Energy Astrophysics,
Code 664, NASA/GSFC, Greenbelt MD 20771}

\begin{abstract}

The restframe mid-ultraviolet spectral region (2000--3200 \AA) is
important in analyzing the stellar populations of the ``red envelope''
systems observed at high redshifts.  Here, we explore the usefulness
of the mid-UV spectral region for determining ages and abundances of
old populations.  We work with a theoretical set of low resolution
spectra and broad-band colors because tests show that these are
presently more realistic than high resolution models.  

A mid-UV to optical/IR wavelength baseline provides good separation of
population components because the main sequence turnoff dominates the
integrated light between 2500 and 4000 \AA.  Mid-UV spectral features
are not sensitive to the dwarf/giant mixture in the population, unlike
those in the optical region.  We find a six magnitude difference in the
mid-UV continuum level (normalized at V) over the metallicity range
$-1.5 < \log Z/Z_{\odot}\ < +0.5$ and a comparable difference (per
unit $\log t$) for ages in the range 4-16 Gyr.  Logarithmic
derivatives of mid-UV colors with respect to age or metal abundance
are 3-10 times larger than for the UBV region.  Most of the spectral
information on old populations therefore resides below 4000 \AA.
Measurement of a single mid-UV color is capable of placing a strong
lower bound on the mean metallicity of an old population.  We
investigate the capability of UBV and mid-UV broad-band colors to
separately determine age and abundance, taking into account precision
in the color measurements.  We find that the mid-UV improves
resolution in $\log t$, $\log Z$ space by about a factor of 3 for a
given observational precision.

Contamination by hot, post-He-flash evolutionary phases can seriously
affect the mid-UV spectra of old populations. A simple estimate shows
that contamination can reach over 80\% in some cases.  However, this
is straightforward to remove as long as far-UV measurements are
available.  We find that extinction should have relatively small
effects on parameters derived for old populations from the mid-UV.

Finally, we show that a 4 Gyr, solar abundance model based on empirical
spectra for nearby stars provides an excellent fit to the mid-UV
spectrum of the Local Group elliptical galaxy M32.  This indicates
that the poorer results obtained from theoretical spectra do
arise from limitations of the synthesis models for individual stars.  

\end{abstract}


\section{Introduction}

\subsection{Mid Ultraviolet Age Dating of Distant Galaxies}

\noindent Recent deep surveys to redshifts beyond $z \sim 0.5$ reveal
galaxies in a wide range of evolutionary states.  Among the
most interesting types are the ``red envelope'' systems 
and the ``extremely red objects'' (ERO's).  These have integrated spectra or
colors which suggest they have experienced little star formation in
the preceding few billions of years 
(Aragon-Salamanca et al.\ 1993;
Kauffmann et al.\ 1996; 
Spinrad et al.\ 1997; 
Ellis et al.\ 1997; 
Zepf 1997; 
Franceschini et al.\ 1998; 
Stanford, Eisenhardt, \& Dickinson 1998;
Dunlop 1999; 
Stiavelli et al.\ 1999; 
Menanteau et al.\ 1999; 
Cimatti et al.\ 1999; 
Daddi, Cimatti, \& Renzini 2000;
Ferreras \& Silk 2000;
Ferreras, Melchiorri, \& Silk 2001;
McCarthy et al.\ 2001; 
Bernardi et al.\ 2002).
The systems which remain in such samples after dust-reddened galaxies
have been purged (e.g.\ Yan \& Thompson 2002) are the least active
galaxies at their epochs in terms of star formation and are
qualitatively similar to quiescent elliptical galaxies at low
redshifts.  They are key probes of the star formation history and
chemical evolution of the universe at epochs considerably earlier than
those at which they are observed.  They can be used, for instance, to
determine when rapid star formation began in the densest
proto-galactic clumps and to what extent it was synchronized in
different environments.  Well-determined maximal ages for galaxies at
different redshifts can be an important constraint on cosmological
parameters (e.g.\ O'Connell 1988, Spinrad et al.\ 1997, Jimenez \&
Loeb 2002).

The restframe ``mid-ultraviolet'' spectral region (2000--3200 \AA) is
especially important in deducing the history of quiescent galaxies from
their integrated light.  At $z \ga 0.5$ this region is redshifted into
the observed optical/near infrared bands, where large ground-based
telescopes can easily access it.  More importantly, the mid-UV should
also be more sensitive directly to the age and metal abundance of the
stellar population than is the restframe optical band spectrum
(O'Connell 1988;
Burstein et al.\ 1988;
MacLaren et al.\ 1988;
Magris \& Bruzual 1993; 
Dorman, O'Connell, \& Rood 1995, hereafter Paper II; 
Bressan, Chiosi, \& Tantalo 1996; 
Yi et al.\ 1997;
Ponder et al.\ 1998;
Smail et al.\ 1998;
Rose \& Deng 1999;
Buson et al.\ 2000;
Ferreras \& Silk 2000;
Ferreras, Scannapieco \& Silk 2002).
The mid-UV light is dominated
by main sequence stars close to the turnoff.  Turnoff light is
much more sensitive to age than is the light of red-giant branch
(RGB) stars, which contribute at least half of the flux at longer
optical and IR wavelengths.  The UV is also more sensitive to metal
line blanketing and therefore abundance than are longer wavelengths.

A telling example of the stakes and controversies involved in studying
the restframe mid-UV of ERO's is the case of the extremely red radio
galaxy LBDS~53W091 at $z =$ 1.552.  Dunlop et al.\ (1996) and Spinrad et
al.\ (1997) estimated from the general shape of its mid-UV energy
distribution and the amplitude of two spectral discontinuities (at
2640 and 2900 \AA) that LBDS~53W091 had an age at its observed epoch
of $\ga 3.5$ Gyr.  They based their estimates on a combination of
empirical stellar spectra and a variety of theoretical spectral
synthesis models.   Such a large age would rule out, for instance, an
Einstein-de Sitter cosmology for $H_0 > 50$ km s$^{-1}$ Mpc$^{-1}$.
It would imply that massive galaxy formation could be completed at
high redshifts ($z > 7$) even in cosmologies with nonzero $\Lambda$
parameters.

Because of its implications, this result has been criticized by several
groups, not on the basis of the difficult observations of LBDS~53W091
itself but rather on the age calibration for the model spectra.
Bruzual \& Magris (1997) used a different set of UV models and fits to
optical colors to estimate an age of 1--2 Gyr, permitting a more recent
collapse time. Heap et al.\ (1998) recalibrated the 2640 \AA\
discontinuity and likewise found a ``young'' age.  Yi et al.\ (2000)
presented fits to the mid-UV spectra and optical-band colors based on
a full, independent grid of models and also obtained a young age.
However, Nolan et al.\ (2001) point out an apparent error in the
evolutionary timescale of the Yi et al.\ models and derive a 3--4 Gyr
age for LBDS~53W091 and a slightly older one for a similar system,
LBDS~53W069.  More recently, Nolan et al.\ (2003) find that, after
calibration using Hubble Space Telescope (HST) observations of two
nearby F stars, their model grid with a range of metallicity and age
places a strong lower limit of 2 Gyr on the age of the two galaxies
and a best fit age $\geq 3$ Gyr.

\subsection{Limitations of Theoretical Models for Mid-UV Spectra}

The disagreements between these various groups suggest that the art of
producing realistic and well-calibrated model spectra for composite
systems in the mid-UV has not yet matured.  The difficulties probably
lie both in the adopted component models (stellar atmospheres and
interiors) and in the calibration of these and conversion to
observables.  These studies all incorporate theoretical spectral
synthesis models for the emergent UV spectra of stellar atmospheres.
Such models are essential to fill out the grid of parameter space
because the available empirical spectral libraries derived from
observations of bright stars (e.g.\ Fanelli et al.\ 1992; Heap et al.\
1998; Rose \& Deng 1999) are largely restricted to metallicities near
the Sun's.

The most serious known weakness is in the synthesis calculations for
cooler stars ($T_e \la 8000$ K).  For abundances appropriate to
elliptical galaxies, a true ``continuum'' is not present in the mid-UV
spectra of stars cooler than 8000 K because of severe line
blanketing.  Relatively few of the hundreds of important absorption
lines have empirical oscillator strengths.  Despite the heroic efforts
of Kurucz (1992), there are quantum-mechanical predictions of
wavelengths and oscillator strengths only for the stronger lines.  In
these calculations, predicted features may be misplaced in wavelength
or may be too strong or too weak.  More reliable empirical
determinations exist for only a much smaller number of UV lines.
Predictions for narrow-band measures of spectral features can also be
a strong function of the spectral resolution of the computations, with
the highest feasible resolution being desired for the best fidelity.

The Kurucz (1992) models for individual stars, which were used in most
of the synthesis attempts cited above, can be directly compared to the
set of low spectral resolution (6 \AA) data for nearby stars obtained
by the International Ultraviolet Explorer satellite (IUE) (e.g.\
Fanelli et al.\ 1992).  The agreement is discouraging.  Lotz et al.\
(2000) find that in only one of seven narrow-band spectral indices
measuring individual strong absorption lines is there good agreement
between the models and the IUE data for stars of spectral types F-G
(spanning the turnoff in old populations).  Similarly, Nolan et al.
(2002) find better general agreement for spectra at low (10-15 \AA)
resolution, although the fits are significantly poorer for an F8 V
star ($T \sim 6100$) than for an F4 V star ($T \sim 6600$), and
deviate systematically in specific wavelength regions (e.g.\
2400--2500 \AA, 2650--2750 \AA).  It is not clear how to interpret
models all of which deviate in some spectral regions in the observed
range even if they can provide good fits to parts of the range.
Formal statistical measures of goodness of fit are suspect under
these circumstances.  

Originally, we had intended this paper to discuss fits to mid-UV galaxy
spectra at 5--10 \AA\ resolution, based on the evolutionary tracks
and isochrone interpolation methods developed in Paper II combined with
Kurucz (1992) atmospheres.  Unfortunately, we encountered the same
difficulties as did Lotz et al.\ (2000):  theoretical narrow-band
indices mostly failed to match empirical values for solar abundance
stars near the turnoff in old populations.  

None of the existing theoretical model sets appear to be of a quality
sufficient to match the observational signal-to-noise level of typical
stellar or galaxy data in narrow bands.  Until theoretical atmosphere
models can demonstrate agreement with the spectral data for individual
stars, we believe it is premature to attempt detailed fits of yet more
complex, composite population models to empirical galaxy spectra.
Fortunately, Heap et al.\ (1998) and Peterson, Dorman \& Rood (2001)
are now attacking the problem of improving the synthetic fluxes based
on iteration against high resolution mid-UV stellar data from IUE and
HST.

However, both we and Lotz et al.\ (2000) found that fits were better for
broader-band measures such as the ``Mg Wide'' index of Fanelli et al.\
(1992), the spectral discontinuities at 2900 and 2640 \AA, and for
broad-band mid-UV colors.  Even though accurate modeling of these
features requires synthesis at high spectral resolution followed by
smoothing over larger bandwidths, it appears that the uncertainties in
line opacities are sufficiently stochastic that they are
reduced in the wider bands.

\subsection{The Mid-UV at Low Resolution}

Although all existing mid-UV model sets are suspect at high resolution,
they give a better approximation to reality at low resolution.
Therefore, we decided to use our models to explore the general
behavior of low-resolution mid-UV spectra and broad-band colors as a
function of age and metal abundance.  Although we calculate the
theoretical spectra at 10 \AA\ resolution, we examine the results only
at lower resolutions. We do not attempt to calibrate the age-abundance
relations in absolute terms, only to investigate relative
dependencies. 

Our main goal is to determine to what extent two well-known
difficulties in studying the integrated light of old populations can be
ameliorated using mid-UV observations.  First, the
integrated photometric properties of populations are approximately
power-law functions of age (O'Connell 1988), i.e.\ $\delta I/I \sim
\textrm{(const)}\, \delta t/t$, where I is any flux or flux ratio
index.  This means that their sensitivity to age declines as age
increases.  The slow photometric evolution of old populations is a
serious obstacle to precise age-dating.  Second, there is a serious
``age-metallicity degeneracy'' (O'Connell 1986, Worthey 1994, Worthey
et al.\ 1995, Bressan et al.\ 1996, and references therein) in that
changes of age or metal abundance can have nearly indistinguishable
effects on colors or line strengths.  Broadband optical colors are
particularly susceptible to both these problems, and this has
contributed to disagreements over the ages and abundances
of old populations found in the literature.  The stronger dependence of
mid-UV light on age and abundance could offer a solution to both
difficulties.

In the next section we discuss our modeling procedures.  We then
describe how long wavelength baselines permit separation of different
population components.  The overall sensitivity of our predicted
energy distributions to age and abundance is discussed in \S 2.3.  In
\S 3 we describe the broad-band colors derived from the energy
distributions and investigate their sensitivity in terms of error
ellipses in the $(t,Z)$ plane.  We also discuss in \S 4 the importance
of corrections for contamination of the mid-UV spectra of old
populations from the high temperature ``UV-upturn'' or ``UVX''
component, which often dominates the light at wavelengths below 2000
\AA.  We investigate extinction effects in \S 5.  Finally, we
illustrate the comparison between a model based on empirical IUE
spectra of stars and the integrated spectrum of M32.

\section{Synthetic Mid-UV Spectra for Old Populations}

\subsection{Method}

This study is based on synthetic spectral energy distributions (SEDs) at
10~\ang\ resolution for single generation populations of a given
abundance.  The SEDs were computed with revised versions of the
evolutionary tracks and the isochrone interpolation routines used
in Dorman, Rood, \& O'Connell (1993, hereafter Paper I) and Paper II.
The evolutionary sequences for this study were computed with the Los
Alamos Opacity Library, rather than the newer OPAL opacities, so that
the results discussed in this paper are consistent with our earlier
calculations.  However, the gross properties of stellar sequences that
affect integrated SEDs are modified only in small quantitative detail
by the use of older physical inputs. The relative lifetimes of
different evolutionary stages, which are the most important factors
affecting SEDs, are not changed, although the mean temperature of a
given isochrone---and thus the age interpretation of a given
observation---may be slightly different.  The effect is almost always
much smaller than the random and systematic errors in
the data and uncertainties in the other modeling assumptions,
especially the stellar atmospheres.  

We use theoretical atmospheres to compute our grid of model SEDs
because no empirical library of stellar spectra has sufficient
metallicity coverage yet.  We generate integrated spectra for each
assumed (age, abundance) pair in our grid by summing model stellar
fluxes taken from the Kurucz (1992) database for each (\teff, $L$)
point along the evolutionary tracks.  To improve the optical colors
over those originally computed for Paper II, models with temperatures
cooler than 3900 K are represented by empirical fluxes selected from
the Gunn \& Stryker (1983)  atlas.  Ages and compositions for the model grid we
illustrate in this paper are tabulated in Table~\ref{ref-calcs}.

The 10 \AA\ resolution we adopted for our model energy distributions
would be entirely adequate for study of broad-band mid-UV colors if the
underlying atomic data for individual spectral lines were accurate.
As noted in the introduction, however, this is not the case.
Accordingly, our UV colors are not expected to be highly accurate, but
they are realistic enough for the general assessment of behavior we
make here.  

We have assumed a Salpeter IMF for all models, $\psi(M) \propto
M^{-2.35}.$   In old stellar populations, the optical/UV light is dominated
by stars near the turnoff mass (0.8-1.0 \msun), and dependence of the 
energy distribution on the IMF is very weak.

\subsection{Spectral Decomposition of the Integrated Light of an Old 
Population}

The contributions from different components of a population can
change dramatically with wavelength.  This is a fundamental feature
bearing on the interpretation of old populations in the UV-optical
region, so it is worth illustrating here.  Similar comparisons for
other model sets, but emphasizing optical wavelengths, have been
published by Buzzoni (1989) and Yi et al.\  (1997), among others.

It is important to divide the population into pre- and post-He-flash
categories.  This is because of the critical effects which mass loss
near the point of the He-flash on the RGB has on the subsequent
evolution.  Since there is no satisfactory deterministic theory for
mass loss, the connection between the pre- and post-flash evolution in
largely ad-hoc (see Willson 2000 and the discussion in Paper II).  As
shorthand, we refer to the pre-He-flash stages as the ``isochrone''
stages.

We have divided the pre-He-flash isochrones into three
sections:  (a) the main sequence for $\teff < 5000$; (b) the turnoff
contribution, which includes all points on the isochrone with $\teff >
5000$ (i.e.\ the upper part of the main sequence and the subgiant
branch, or SGB); and (c) the red giant branch (RGB) itself from the
point where $\teff < 5000$.  This decomposition allows us to see
separately the contribution to the integrated light from hotter stars
of the earlier evolutionary phases, whether they lie on the main
sequence or the subgiant branch.  

\begin{figure}[t]
\plotone{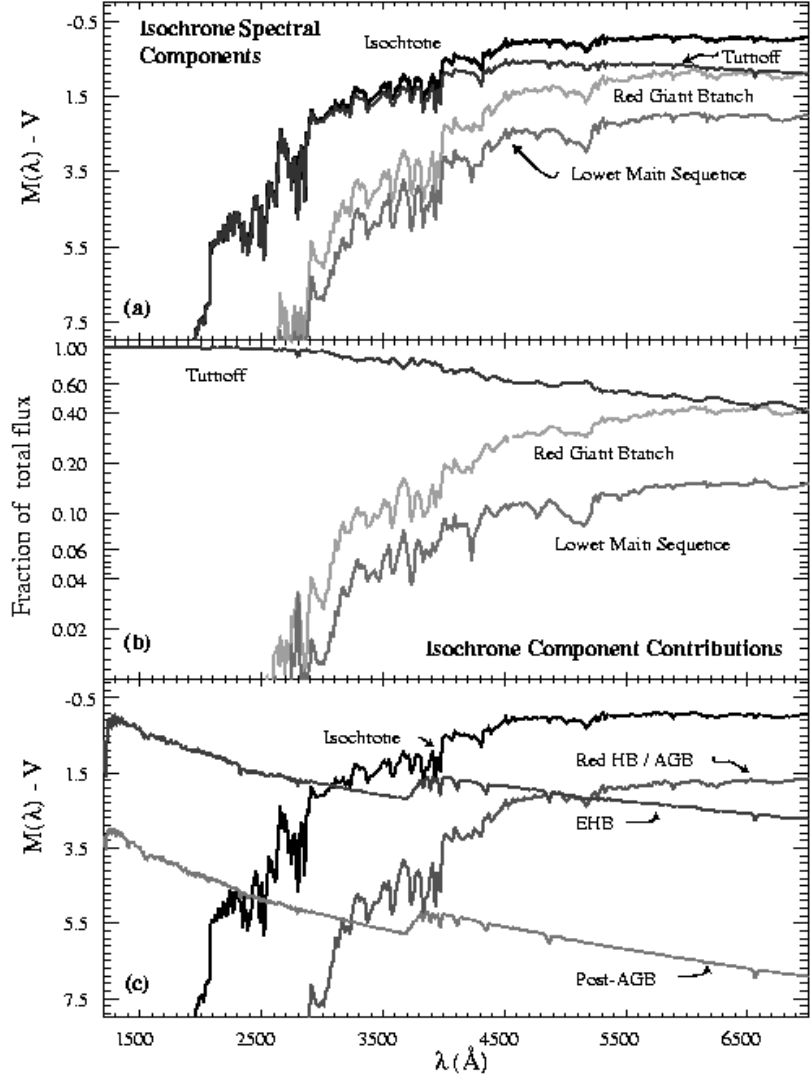}
\caption {{ \small \protect\label{fig:spec-decomp}{\singlespace
Decomposition of the spectral energy distribution of an old stellar
population (10 Gyr age, solar metallicity) into its various components.
Panel (a) shows the decomposition of the pre-He flash sequence (the
``isochrone'' component) into three parts: the Red-Giant Branch
($\protect\teff < 5000
\kelvin, L > L_{TO}$), the turnoff region ($\protect\teff > 5000 \kelvin$),
and the lower main sequence ($\protect\teff < 5000
\kelvin, L > L_{TO}$). Spectra are plotted in magnitudes normalized
to the $V$--band.
Panel (b) shows, on a logarithmic scale, the contributions of each constituent to
the isochrone SED as a function of wavelength. Panel (c) compares the
isochrone component to the maximum possible SED contributions of the 
three main advanced evolutionary stages (Red HB/AGB, Post-AGB, and
Extreme HB).  These are shown as  
though all evolving stars passed through that particular phase.  
}} }

\end{figure}

Figure~\ref{fig:spec-decomp} illustrates the relative contributions to
the integrated SED of the various components.  We take
the models at 10 Gyr for solar metallicity as an example.  We plot the
wavelength range  \wrange{1000}{7000} to contrast the behavior of the
SED in the UV with the optical.  Panel (a)  shows the predicted SED of
the  isochrone and its components, plotted as 
$(M_{\lambda} -V) = -2.5 \log [ F_{\lambda}(\lambda)/\langle F_{\lambda} \rangle_V]$,
where 
$F_{\lambda}$ is the flux per unit wavelength and
$\langle F_{\lambda} \rangle_V$ is the mean flux per unit wavelength
in the V filter band (see \S 3).  
Panel (b) shows the three components of the
isochrone SED plotted as fractions of the total isochrone flux.  

Panel (c) compares the SEDs of other potential components of an old
stellar population. The post-He-flash, advanced evolutionary stages
include the red horizontal branch and the asymptotic giant branch
(RHB$+$AGB), the extreme horizontal branch (EHB), and the
post-asymptotic giant branch (P-AGB). The total population in these
advanced phases has been scaled by the evolutionary flux $\dot n_0(t)$
predicted from the isochrone (see Paper II), but the spectra are
illustrated here as if all the stars reaching the He-flash continued
into that particular channel.  The plotted spectra therefore represent upper
bounds to the contribution from each phase.  

The figure illustrates several points:

\begin{enumerate}

\item While the contribution from the turnoff stars and the giants is
each about 50\% in the V band, the turnoff accounts for most of the
flux in panels (a) and (b) shortward of 4000 \AA, reaching $\sim 90\%$
at 3000\AA.  This dramatic change in component dominance allows
observations with a long wavelength baseline to ``dissect'' the
stellar population.  

\item The RHB$+$AGB contribution is dominated by cool AGB stars.  Its
maximum contribution is about 1 mag fainter than that of the
RGB stars, consistent with the shorter lifetimes of the more
advanced phases.  The amount of cool RHB$+$AGB light depends on the
amount of RGB mass-loss and in principle could vanish altogether if
mass-loss were large enough.  Thus, RGB mass-loss can affect
{\it both} the UV and the optical-IR SEDs. 

\item Feature strengths in the optical region ($>4000$\AA) are
influenced by both giants and dwarfs, meaning that detailed modeling
of the surface gravity mixture is necessary to interpret them.  By
contrast, the cool RGB, RHB, and AGB objects do not contribute
significantly to the mid-UV light, which means that a much ``cleaner''
interpretation of the population structure is possible there.  

\item The hot EHB and P-AGB components are responsible for the spectral
upturns or ``UVX'' at wavelengths below 2000 \AA.  The EHB is defined
to be those horizontal branch stars with such small envelope masses
(typically $\la 0.05$ \msun) that they do not evolve back up the AGB
(Greggio \& Renzini 1990, Paper II).  If all post-RGB stars passed
through the EHB, they would give rise to a UVX which is nearly 2
magnitudes brighter than the largest observed to date (Paper II and
O'Connell 1999).  Evidently, in the available sample of galaxies, only
a minority of the post-RGB objects pass through this phase.  The
maximum P-AGB contribution is much smaller than the EHB maximum owing
to the very short lifetimes of these objects (Greggio
\& Renzini 1990; Paper II).

\item Depending on the fraction of post-He-flash stars in each channel,
either the EHB or P-AGB components could significantly affect light in
the 2000--3500 \AA\ region.  This UVX contamination complicates the
interpretation of the mid-UV light but can be dealt with
straightforwardly in practice (see \S 4).

\end{enumerate}

In the remainder of this paper, we have normalized our model energy
distributions in the V-band so that we can conveniently compare the
behavior of the mid-UV and optical bands.  The use of V-band
normalization means that our results are not independent of the
influence of the giant branch.  They are partially subject to the
various uncertainties which have been revealed in optical-band studies
(e.g.\ Charlot et al.\ 1996).  A surprising difficulty which has
recently emerged is that the giant branch luminosity function is
significantly underestimated by the theoretical models (Schiavon et
al.\ 2002).  In practice, one could avoid such complications while
retaining the advantages of the mid-UV by confining one's attention
strictly to the shorter wavelengths where the main sequence dominates,
using the 3400 \AA\ region, for instance, as a normalization.

\subsection{Sensitivity of Predicted Mid-UV Energy Distributions to Age and Abundance
\label{sec:agemetal} }

In this section we illustrate the general sensitivity of the predicted
mid-UV SEDs to metallicity and age, excluding complications such as UVX
contamination. 

\begin{figure}[t]
\plotone{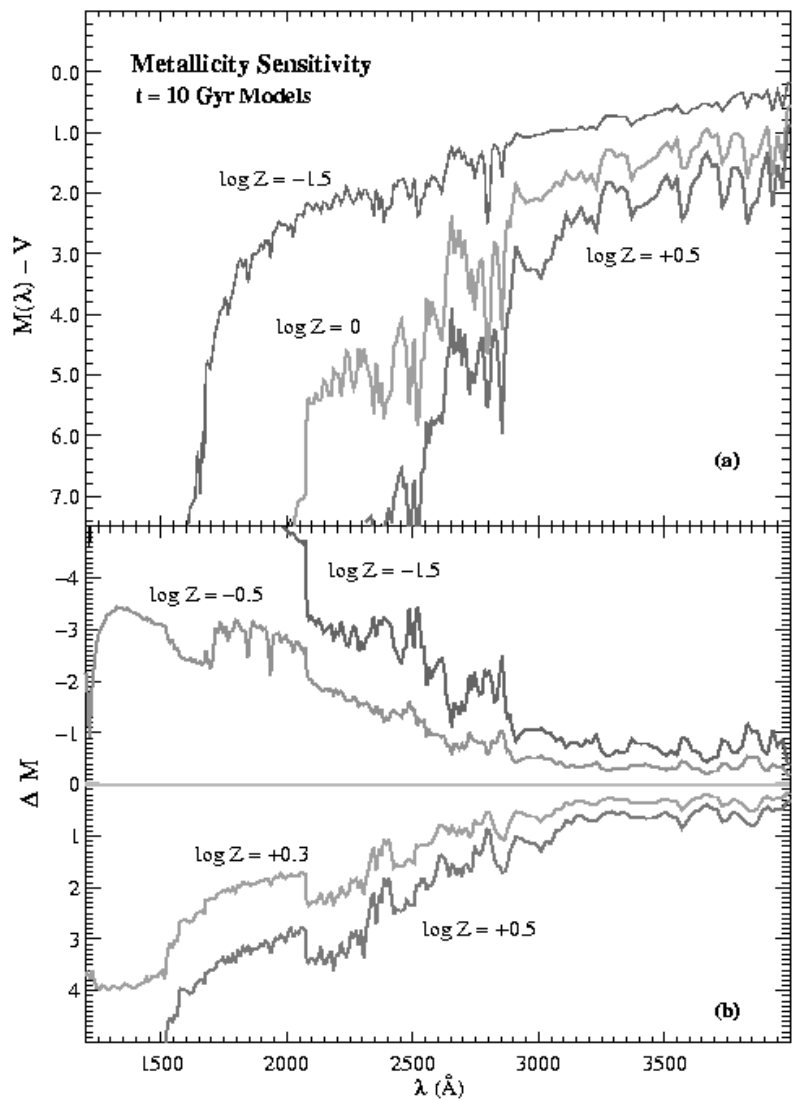}
\vspace{10pt}
\caption{\protect\label{iso-metals} 
The sensitivity of model isochrone SEDs to metallicity at fixed age.
Panel (a) shows 10 Gyr-old populations for metallicities with a factor
of 100 range bracketing solar abundance.  Metallicity labels give
values with respect to solar abundance.  The spectra are plotted in
magnitudes normalized to the $V$--band. The huge increase in UV
continuum blanketing for higher metallicities is evident.  Panel (b)
shows difference spectra with respect to the solar metallicity SED,
again at 10 Gyr.} 
\end{figure}

Figure~\ref{iso-metals} shows the integrated spectra and difference
spectra for the ``isochrones'' (i.e.\ pre-He-flash evolution) with 
metallicities $-1.5 < \log Z/Z_{\odot} < +0.5$ at 10~Gyr for the wavelength range
\wrange{1200}{4000}.  The spectra in Fig.~2 are all normalized at the
V-band.  The large range in flux level for $\lambda < 3000$ \AA\
graphically illustrates the extreme sensitivity of the mid-UV to metal
abundance, which amounts to $\pm 3$ magnitudes in the cases plotted.
The spectra at $ \log Z/Z_{\odot} = -0.5$ and 0 have a strong break at
2100~\AA\ whereas at lower metallicity there is significant flux
shortward of this limit.  Blanketing in the super-solar metallicity
spectra produces a sharp decrease in the continuum shortward of
2600~\ang.  Even over the range $ -0.5 < \log Z/Z_{\odot} < +0.3,$
arguably of most interest for studies of galaxies, Fig.~2(b) shows
that the continuum level at a fixed age of 10 Gyr differs by about 1
mag at 3500~\AA, rising to a $\sim 3$ mag separation at 2400~\AA.

Fanelli et al.\ (1992) explored the metallicity and temperature
sensitivity of mid-UV features in stars observed by IUE.  In general,
stellar temperature is more important than metallicity.  The effects
seen in the predicted integrated spectra of Fig.~\ref{iso-metals} are
therefore mainly driven by the influence of metallicity on the
temperature of the isochrones, rather than by the influence of
metallicity directly on line strengths.  The contrast of individual UV
spectral features with the local ``continuum,'' and therefore the
apparent strength of the features, can be strongly affected by the
very large blanketing in the continuum.  Fanelli et al.\ and Rose \&
Deng (1999) found that the strength of some UV features (e.g.\ the
Mg~{\textsc{ii}} doublet) is actually larger in more metal poor stars,
mainly because of diminished continuum blanketing.  Some of these
effects are apparent on close inspection of Fig.~2.  

The rapid decline below 3000 \AA\ in the flux of the more metal rich
models in Fig.~2 clearly implies practical difficulties in obtaining
spectra or photometry of such systems.  It could  contribute to
selection effects in the samples of objects detected at high redshifts.

\begin{figure}[t]
\plotone{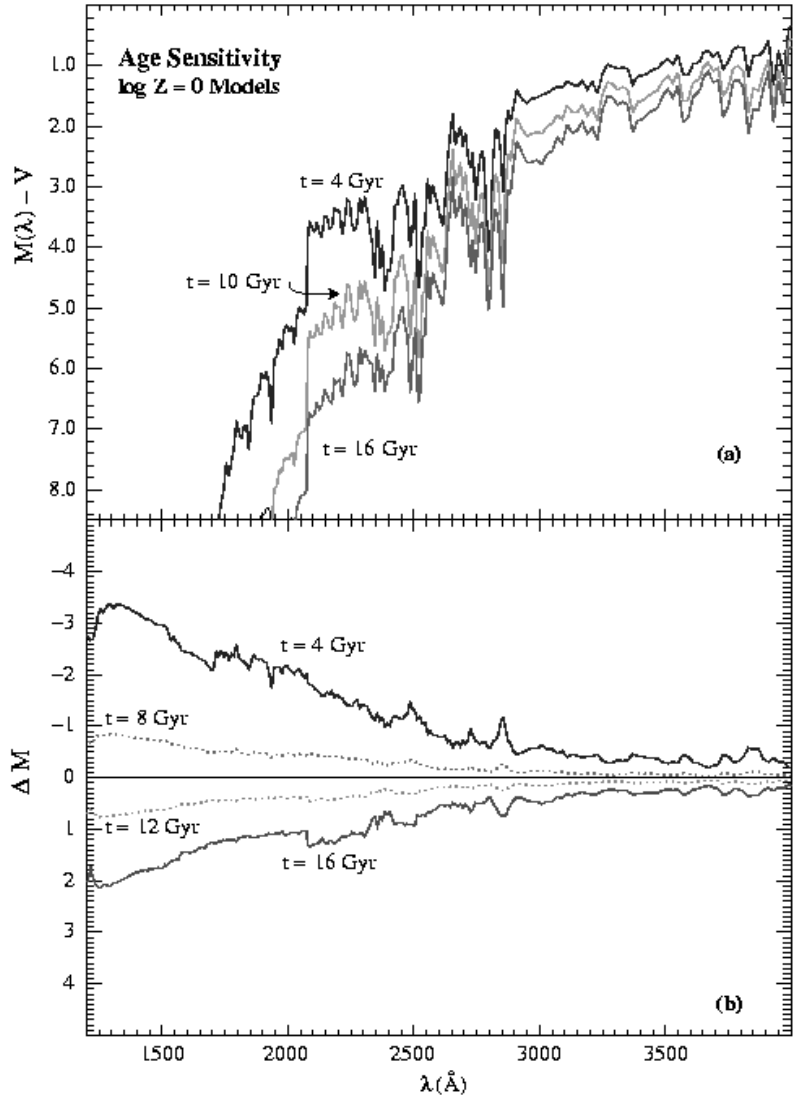}
\vspace{10pt}
\caption{{\small\protect\label{iso-ages} {\singlespace The sensitivity
of model spectra of solar abundance to age. Spectra in panel (a) are
shown in magnitudes normalized to the $V$--band.  Panel (b) shows
difference spectra with respect to the model for 10 Gyr.  }}}
\end{figure}

In Figure~\ref{iso-ages} we illustrate age effects on the spectra of
isochrones for solar abundance in a format like that of Fig.~2.
Difference spectra in the lower panel are relative to the 10 Gyr
model.  Qualitatively, the difference spectra resemble those of the
metallicity difference spectra in Fig.~2.  Their amplitudes are smaller
because the total logarithmic range considered is smaller (0.6 dex in
age compared to 2.0 dex in metallicity).  However, the amplitude per
unit change in $\log t$ is comparable.  Their shapes also differ in
detail, and the region 2500-3000\AA\ looks promising for age-abundance
discrimination.

Difference amplitudes such as those in Figures 2 and 3 become much
smaller at wavelengths above 4500 \AA.
(Quantitative values for partial derivatives of color are discussed in
the next section.)  To put this another way, most of the spectral
information on old stellar populations resides below 4000 \AA.  

The sensitivity of the mid-UV to population parameters is obviously
large, but is it sufficient to distinguish the effects of metallicity
from those of age?  Taking the models at face value, the answer is
yes, as illustrated in Figure 4.  Here we plot the difference between
the spectra of a younger, metal rich model and an older, solar
abundance model.  This particular pair (6 Gyr, 3$Z_{\odot}$),(16 Gyr,
$Z_{\odot}$) was chosen because it was found by Worthey (1994) to be
nearly indistinguishable at optical wavelengths.  From Fig.~4, our
models confirm Worthey's (1994) conclusion for the spectral range
4500--6000 \AA: apart from the strong Na I doublet at 5890 \AA, the
pair of SEDs is remarkably similar.  But large differences between the
two models appear at shorter wavelengths.  The transition to higher
sensitivity occurs near 4000 \AA, where the strong decline in cool
stellar energy distributions caused by both temperature and line
blanketing effects begins.  Both the continuum and strong absorption
features (almost all of which are blends of different species) are
useful age-abundance discriminants in the 3000--4000 \AA\ region.  Data
with typical precisions of 0.05 mag for 3000--4000 \AA\ can readily
distinguish this pair of models.  Very large differences appear at
3100 \AA\ and increase sharply to shorter wavelengths.

The fidelity of the higher abundance model in this comparison is open
to doubt, so Fig.~4 should be viewed only as broadly indicative of
spectral behavior.  More than anything else, this diagram emphasizes
the importance of the largest practical wavelength coverage in
studying old populations.

\section{Mid-UV/Optical Broad-Band Colors \label{sec:muv-opt} }

Now we consider the behavior of broad-band mid-UV colors computed from
the full SEDs of our models.  We use the colors primarily as tracers to
characterize the general response of the mid-UV spectral region to
population parameters.  We anticipate that narrow-band spectral
measures, once they are properly calibrated by accurate models, would
yield yet better performance.  However, broad-band colors are useful
in practice for several reasons:  (a) they are more easily and
reliably modeled than are individual spectral features; (b) it is
easier to achieve good observational precision in broader bands, an
important consideration for high redshift galaxies; and (c) they are
not strongly affected by complicating factors such as velocity
dispersion or emission features.  Here, we discuss the ability of
combined mid-UV and optical colors to determine age and abundance.

Chiosi, Vallenari, \& Bressan (1997) and Yi et al.\ (1999) discussed
the usefulness of broad-band UV colors for the study of old
populations, but they emphasized the far-UV region below 2000 \AA\ and
employed the UVX component as a primary dating tool.  The UVX is
certainly a potentially good index to galaxy ages and abundances, but
its interpretation is subject to large prevailing uncertainties in
mass loss physics (O'Connell 1999, Willson 2000).  Despite the
difficulties with model atmospheres, interpretation of the mid-UV is
currently on surer physical ground.  Chiosi et al.\ (1997) and Yi,
Demarque, \& Oemler (1995) also emphasize practical difficulties with
red leaks in far-UV filters, such as those on the Hubble Space
Telescope.  Red leaks are less serious for the mid-UV range.

We have computed synthetic colors from our model SEDs by integrating
filter transmission curves over them.  The colors are given in
Table~\ref{tab:color}(a--e) as a function of age and metallicity. We
have used the HST/WFPC2 filters F218W, F255W, and F300W to illustrate
the behavior of broadband mid-UV colors. These filters have central
wavelengths and effective widths defined in Biretta et al.\ (1995) as follows:
F218W: $\lambda_{\textrm{\scriptsize peak}} = 2091 \,
\ang,$ $\Delta \lambda = 356 \ang;$  F255W:
$\lambda_\textrm{\scriptsize peak} = 2483 \, \ang,$ $\Delta \lambda =
408 \,\ang;$ F300W:  $\lambda_{\textrm{\scriptsize peak}} = 2760 \,
\ang,$ $\Delta \lambda = 728 \ang.$ HST ultraviolet magnitudes are on
the STMAG system, derived from monochromatic fluxes per unit
wavelength $m(\lambda) = -2.5\log (F_\lambda) - 21.1$ as described in
Biretta et al.\ (1996).  We used
the Johnson $V$ band filter response as given by Bessell (1990), and
we computed for comparison the Johnson $(U-V)$ and $(B-V)$ colors, which have
been normalized to 0 for the empirical spectrum of Vega given by
Hayes \& Lathan (1975). This differs slightly (a few 0.01 mag) from the
standard UBV system.  As shorthand, we refer to the colors produced
from the F218W, F255W, and F300W filters as $(22-V)$, $(25-V)$, and
$(30-V)$, respectively.  

Note that because of the precipitous decrease of flux in the SEDs of
old populations below 3000 \AA, predicted colors are sensitive to the
exact center, shape, and width of the filters used. A small difference
in the sensitivity of the red edge of the mid-UV filter transmission
curves can change the UV colors by $\sim 10$\%.

In Figures 5 and 6 we plot the various colors as a function of $(B-V)$.
The overlap of the $(U-V,B-V)$ colors in Fig.~5 is symptomatic of the
age/abundance degeneracy.  The separation between the model sequences
at different $Z$ in Fig.~6 shows the extent to which the degeneracy
has been lifted in the mid-UV.  The separation is considerably greater
in $(22-V)$ and $(25-V)$ than in $(30-V)$ and $(U-V).$ The separation
between the model pair discussed in \S 2.3 is also evident here.

Applying the plotted models to existing UV data for E galaxies, we
find that some ranges of metallicity are excluded altogether by a
single mid-UV broadband observation (as discussed in Paper II).  For
example, $(25-V) > 3.3$ for most of the luminous elliptical sample in
Table 1 of Paper II. This excludes models with $\log Z/Z_{\odot} <
-0.3$.  

Another way to look at sensitivity is to consider the partial
derivatives $\partial\, {\rm Color} / \partial \log k$, where $k$ is
either age or fractional metal abundance.  Such logarithmic
derivatives tend to be roughly constant for ages above 0.5 Gyr (e.g.\
O'Connell 1988, Bruzual \& Charlot 1993, Girardi 2000).  In Table 3 we
have evaluated derivatives for the five computed colors in the
vicinity of a fiducial model for $t = 10\,$ Gyr, $Z = Z_{\odot}$.  We
see from the table that the derivatives for mid-UV colors are 3--10
times larger than for optical bands.  The derivatives increase rapidly
below 3000 \AA.  This demonstrates quantitatively the improved
sensitivity of the shorter wavelengths to age and abundance in old
populations.  

Nonetheless, the differences between the tabulated $t$ and $Z$
derivatives are not as large as might have been hoped.  Considered
geometrically, the angular separation between surfaces of constant $t$
and constant $Z$ in mid-UV broad-band color space is not much
different than in the optical bands.  However, the higher mid-UV
sensitivity means that a given $t,Z$ change will produce a larger
amplitude effect.  The age-metallicity degeneracy cannot be ignored
even in the mid-UV.  A multi-wavelength approach is clearly required.

\subsection{Observational Discrimination Between Age and Metallicity
in Broad Band UV Colors \label{sec:amd}}

Figures 4 and 6 indicate that age/abundance ambiguities ought to be
significantly reduced in the mid-UV.  However, UV fluxes cannot yet be
measured with the same precision as optical fluxes due to remaining
calibration problems and the faintness of galaxies in the UV.
Therefore, the issue is not simply the sensitivity of the colors to
population parameters, as measured by partial derivatives or diagrams
like Figs.~2-6, but rather the sensitivity compared to
typical observational uncertainties in the various bands (O'Connell
1988).  In this section we compare the sensitivities of integrated
mid-UV and optical colors to age and abundance while taking
observational uncertainties into account.  We consider here only the
case of a simple stellar population: i.e.\ a single generation with a
uniform metal abundance.  

We denote the broad-band colors derived from the spectral energy
distribution for a population of given age, $t$, and abundance, $Z$
by $C(t,Z)$.  The range of models that produce a value of $C$ within a
given observational uncertainty, \sigobs, of a given color, $C_0$, is
the set of $(t,Z)$ such that

\begin{equation}
\label{sig}
\Delta C = \Big| C(t,Z) - C_0 \Big| \le \sigobs.
\end{equation}

For a set of $n$ observed colors $\{C_i\},$ all of which have
comparable observational uncertainties, the range of models consistent
with the data is the error ellipse in the $(t,Z)$ plane defined by the
root mean square (rms) differences in colors between theory and
observation is determined by 

\begin{equation}
\label{r}
R =  \sqrt{ {{1}\over {n}} \sum_{i=1}^{n}\left[\Delta C_i \right]^2 } \le
\sigobs.
\end{equation} 

The $\Delta C$'s are related to the basic population parameters $t$ and
$Z$ as follows: 

\begin{equation}
\label{final}
\Delta C = \Bigg | {\Bigg({\partial C \over \partial \log t}\Bigg) }_Z d \log t  +
{\Bigg({\partial C \over \partial \log Z}\Bigg) }_t d \log Z \Bigg |.
\end{equation}

Worthey (1994) has presented a similar discussion of index sensitivity,
though without explicit inclusion of observational uncertainties.  He
defined ``metallicity sensitivity'' indices in terms of the ratios
of the partial derivatives in this expression
for various combinations of line strengths in the optical band.  

Figure~\ref{ellipse} shows sample error ellipse contours of $R$ in the
$Z,t$ plane as defined in equation~(\ref{r}). To construct this plot,
we have selected a fiducial model (10 Gyr, $Z = Z_{\odot}$, indicated by
the filled circle) and determined its rms separation from other models
in various colors.  In panel (a) the contours are derived using only $(U-V),
(B-V),$  while panel (b) is constructed for mid-UV plus optical colors.

In both diagrams there are obvious ``valleys of degeneracy'' running
diagonally.  Within these valleys an older, more metal poor model can
be found which yields the same colors (for a given level of
observational precision) as a younger, more metal rich model, so it is
not possible to separate $t$ and $Z$ unambiguously.  These are the
analogues in broad-band colors of the ``3/2'' degeneracy relation for
optical absorption-line features discussed by Worthey (1994).  However,
the contours are much broader for a given observational precision in
the optical than in the UV bands.  The innermost contour plotted in
both panels is 0.03 mag, a precision that can be routinely achieved
for nearby galaxies in the optical but not currently in the UV.
However precisions in the range 0.05--0.1 mag are possible in the
mid-UV.  The diagram shows that the mid-UV colors yield uncertainties
for $\sigobs = 0.3$ comparable to those of optical colors at $\sigobs
= 0.1$.  

One can therefore say that broad-band mid-UV colors improve resolution
in $\log t,\log Z$ space by about a factor of 3 for a given
observational precision.  Comparable or better improvement is expected
if one uses narrower bands to measure the highly structured
mid-UV spectrum, as Figures 2, 3, and 4 suggest.  This advantage is
such that it is certainly desirable to push for better quality in both
observations and modeling of the mid-UV.  

Note that we are addressing only the effects of observational error on
derivation of population parameters.  A different kind of uncertainty
was discussed by Charlot et al.\ (1996):  namely, the systematic
uncertainties inherent in the models themselves, which would be
reflected in changes in the contours in these diagrams with different
authors.  Needless to say, both factors contribute to the uncertainty
of the results from population synthesis.

\section{UVX Contamination of Mid-UV Galaxy Spectra \label{sec:uvx}}

The mid-UV spectra of old populations cannot be considered in
isolation from the UVX phenomenon.  Old populations contain low-mass
stars in advanced (post-He-flash) evolutionary phases with $T_e \ga
20000\,$ K, which produce copious UV flux (Greggio \& Renzini 1990;
O'Connell 1999, and references therein).  Although their contribution
is negligible at visible wavelengths, these UVX objects can contribute
significantly to the mid-UV spectral range of interest here.  The
difficulty is that we do not yet have a firm enough understanding of
these evolutionary phases to predict their contribution from first
principles.  Fortunately, the UVX affects the mid-UV spectrum only by
contributing an essentially featureless continuum.  Thus, if
appropriate far-UV observations of the same object are available,
this can be removed empirically, and the spectrum of the cool
components can be extracted with good precision.

In the simplest approach to removing the UVX contamination, one can
take advantage of the facts that the SEDs of the hot UVX objects are
smooth in the mid-UV wavelength range and do not depend strongly on
UVX temperature while the other components in E galaxy populations
contribute little to the far-UV flux ($< 2000\,$\AA).  No assumptions
need be made about the characteristics of the cool population other
than that its far-UV contribution can be ignored.  One can use either
broad-band colors or narrow band fluxes at selected wavelengths.

Here, we use two broad-band colors to estimate the magnitude of UVX
contamination.  In equation form, if we observe galaxy broad-band
colors $(15-V)$ and $(25-V)$ corresponding to UV/optical flux ratios
$\alpha$ and $\beta$, then

\begin{eqnarray}
\label{eqn:uvxcomb}
\alpha  & \equiv &  \fxv_{\textrm{\scriptsize obs}}  =  { {\fxvh + \fxvc} \over {\fvvh + \fvvc} } \nonumber\\
\beta    & \equiv  & \fxxv_{\textrm{\scriptsize obs}}  =    {  {\fxxvh + \fxxvc} \over {\fvvh + \fvvc} },
\end{eqnarray}

\noindent where $F^i_j$ denotes the flux contributed by the $i^{th}$
component at wavelength $j$.
Defining $p$ to be the fractional contribution of the hot component to
the V-band light, and using the fact that the sum of the fractional
contributions of the hot and cool components is 1.0, we have

\begin{eqnarray}
\label{eqn:uvxsol}
\alpha  & = & p\, \fxvvh  + (1-p)\, \fxvvc \nonumber\\
\beta   & = & p\, \fxxvvh + (1-p)\, \fxxvvc.
\end{eqnarray}

\noindent If we assume that $\fxvvc \approx 0$, then for any
photometric properties adopted for the UVX component, i.e.\
for a given pair of \fxvvh and \fxxvvh values,  we immediately obtain an
estimate for $p$.  This also determines the value of $F_{2500}/F_V$ of
the cool component.  The fractional contribution of the hot component
to the light at 2500 \AA\ is $p^{\prime} = (p/\beta)\fxxvvh$.  

The spectrum of a star with $T_e = 24000\,$K is a good approximation
to the UVX energy distribution of luminous E galaxies (e.g.\ Brown et
al.\ 1997).  This has $\fxvvh = 33.0 $ and $\fxxvvh = 10.1$.  In 
Table~4 we show the results of applying these values to the
extinction-corrected, observed broad-band nuclear colors of four
bright galaxies (taken from Paper II).   M60 and
M89 have large UV-upturns, the center of M31 has a moderate upturn,
and M32 has a weak upturn.  The column headed $p$ shows the
contribution of the UVX component to the $V$-band light, while the
column headed $p^{\prime}$ gives the UVX contribution at 2500 \AA.
For M32, the UVX contamination in the mid-UV is minor (only 10\%), but
in the two strong-UVX sources, the UVX light is actually dominant
(60-80\%).  The column headed $(25 - V)_{\rm COOL}$ contains the
estimated color in the absence of the UVX contamination; this can
obviously be very different from the observed color.  Because of the
steep rise of the cool star energy distributions to longer
wavelengths, UVX contamination will decline rapidly longward of 2900 \AA.

The estimated mid-UV contamination in these examples is comparable to
those inferred by Burstein et al.\ (1988) and Ponder et al.\ (1998).
Rose \& Deng (1999) also derived a good fit to narrow-band mid-UV
features in M32 using a 10\% UVX contamination.  For the strong upturn
cases, contamination strongly affects the mid-UV and produces a
significantly bluer and weaker-lined appearance than the intrinsic
cool-star spectrum.  Correction for the contamination is obviously
essential before the mid-UV spectra can be analyzed for age and
abundances.  (We do not attempt to interpret the $(25 - V)_{\rm COOL}$
values using our models because of the various calibration
uncertainties discussed in \S 2.1.)  The UVX itself should
evolve with time and is not expected to be present in systems younger
than a few Gyr (e.g.\ Bressan et al.\ 1996; Yi et al.\ 1997).  

The confidence with which UVX contamination can be removed depends on
the nature of the UV data at hand.  If only a far-UV mean flux is
available, then one must assume a surface temperature and metallicity
for the UVX component.  Uncertainties with such assumptions are likely
to be $\pm 3000$ K in temperature and $\pm 0.3$ dex in $\log Z$.
According to the Kurucz (1992) models, the corresponding changes in
the $(15-25)$ colors of hot stars are $\pm 0.08$ mags and $\pm 0.05$
mags, respectively.  For a typical galaxy in Table~4, $\delta (25 -
V)_{\rm COOL} \sim 1.6\times \delta(15-25)$.  With more data points,
the far-UV spectral slope can be determined and the $(15-25)$ color of
the hot component estimated directly. 

In the case of high quality mid-UV spectra, it would be possible to
obtain a simultaneous solution for the hot and cool components using
optimizing synthesis techniques.  That approach would take advantage
of the fact that the UVX spectrum is expected to be smooth in the
mid-UV.  But even here, it would be desirable to have far-UV
observations available.

\section{Extinction Effects}

A potential complication for analysis of mid-UV light is interstellar
extinction, which is larger there than at optical wavelengths.  In
this section we discuss the likely impact of extinction effects
on the mid-UV.

The fourth column of Table 3 lists the color changes expected for a
color excess of $E(B-V) = 1.0$ based on the UV reddening law of
Cardelli, Clayton, \& Mathis (1989).  The effects of a given amount
of dust in the line of sight on the mid-UV colors are 3-6 times larger
than at optical wavelengths.  However, the net impact of an error in
the extinction correction on a derived population parameter would be in
proportion to the ratio of the derivative of color with respect to
extinction to the derivative with respect to the parameter.  Because
the mid-UV sensitivity to age and abundance is so high, we find that
extinction-driven errors in population parameters derived from
broad-band colors are actually smaller in general in the mid-UV than
in the optical bands.

Internal extinction is generally very small in most nearby elliptical
galaxies.  Far-UV observations by the Hopkins Ultraviolet Telescope
(e.g.\ Ferguson \& Davidsen 1993, Brown et al.\ 1997) provided excellent
limits on internal extinction in a small sample of early-type
systems.  These showed that the slope of the far-UV spectral rise in
the galaxies is nearly at the maximum encountered among hot stars
(Dean \& Bruhweiler 1985, Fanelli et al.\ 1992), leaving little room for
internal interstellar reddening in excess of a few 0.01 mags in $E(B-V)$.
Although high resolution HST images reveal dust lanes or filamentary
structures in many early-type nuclei (e.g.\ Jaffe et al.\ 1994, Tran
et al.\ 2001), these are on such a small scale that they usually do
not importantly affect integrated light measures with apertures
larger than a few arc-sec.  Of course, the less mature systems
observed at large lookback times might exhibit larger extinction
effects than these local counterparts.  

At higher spectral resolutions, where line strengths and spectral
discontinuities can be well measured, it is possible to determine
extinction reliably using optimizing spectral synthesis techniques
(e.g.\ O'Connell 1980; Fanelli, O'Connell, \& Thuan 1988).  The method
works because the absorption line strengths and sharp continuum breaks
are sensitive only to the stellar population mixture while the
continuum slopes over wider wavelength intervals also respond to
extinction.  For optical spectra, the technique yields uncertainties
in $E(B-V)$ of a few 0.01 mags, and comparable performance should be
possible in the mid-UV given good quality data.  Spinrad et al.\
(1997) based their age-dating for a galaxy at $z = 1.55$ in part on
the spectral breaks at 2640 and 2900 \AA, which are essentially
reddening-independent.  

We have also examined the extent to which reddening will affect
correction for the UVX contamination.  For color excesses (or
reddening correction errors) up to $E(B-V) \sim 0.15$, we find that
the effects on the derived $(25 - V)_{\rm COOL}$ (see \S 4 and Table
4) are small, of order 0.05 mag.  The reason is that excess reddening
reduces the far-UV flux relative to the mid-UV but simultaneously
causes the far-UV energy distribution to look cooler, thereby
partially compensating the estimated contamination fraction.  

We conclude that, although extinction corrections should be part of
any analysis of mid-UV spectra, extinction is not a major
impediment to using the mid-UV to study old populations.

\section{An Empirical Library Model for M32}

The empirical mid-UV spectral libraries, based on bright stars, offer
good coverage of the temperature and gravity grid for near-solar
metallicity.  Although these may be of limited usefulness for
interpreting the spectra of luminous E galaxies, which tend to be more
metal-rich, they should be satisfactory for systems which have
solar-like abundances.  The most important of these is the Local Group
elliptical M32.  This is well established to have a mean metallicity
near solar (e.g.\ O'Connell 1980, Rose 1994, Worthey 1998, Rose \&
Deng 1999, Vazdekis
\& Arimoto 1999, Trager et al.\ 2000, del Burgo et al.\ 2001) and
is bright enough that mid-UV data with good signal-to-noise is
available.  It is interesting to explore the extent to which the
available empirical libraries are capable of providing good fits to the
spectra of galaxies like M32.

Here, we show a fit to the 6 \AA\ resolution IUE spectrum of the
center of M32 obtained by Burstein et al.\ (1988; data kindly provided
by D. Burstein).  The spectrum was corrected for foreground extinction
using the Savage \& Mathis (1979) law and adopting $E(B - V) = 0.077$
(Burstein \& Heiles 1984).  Correction for UVX contamination was based
on the results of the previous section and employed a synthetic UVX
SED matching the resolution of the data.  Since the contamination is
small, the resulting spectrum does not depend strongly on the
correction.  The model illustrated (Fanelli 1991) has solar abundance
and a turnoff age of 4 Gyr, similar to that derived in the various
optical-band studies cited above.  The stellar spectra combined in the
model were taken from the library of mean IUE spectra for stellar
groups published by Fanelli et al.\ (1992).  The assumed component
contributions by each group are listed in Table 5.  

The correspondence between the model and the data shown in Figure 8 is
excellent.  The model provides a good fit to the pseudo-continuum
across the entire range illustrated (2100-3200 \AA), showing slightly
larger residuals below 2400 \AA, where the signal-to-noise of both the
galaxy and input stellar data becomes poorer.  It also fits all the
strong absorption features above 2500 \AA.  (Rose and Deng 1999 had
already shown good agreement between solar abundance stars and M32's
spectrum for these features.)  There was no attempt to optimize the
fit (e.g.\ by varying the mixture of components) other than adding a
zero point offset of $-0.22$ magnitudes to the corrected data.  This
offset is probably within the uncertainty in matching the IUE and
optical photometric apertures.  

Figure 8 shows that empirical mid-UV stellar spectra can provide
very good fits to galaxy spectra where metallicities match.  This argues
that the poorer results obtained with theoretical spectra do arise from
limitations of the synthesis models for individual stars.

\section{Summary}

Good tools for the analysis of the mid-UV spectral range (2000-3200
\AA\ in the rest frame) are needed to interpret stellar populations in
old galaxies at high redshift.  None of the existing theoretical model
sets appear to be of a quality sufficient to match the observational
signal-to-noise level of typical mid-UV data in narrow bands, mainly
because of the difficulty of producing high fidelity synthetic spectra
for metal rich stellar atmospheres.  In this paper, we have used models
at low spectral resolution, which are expected to be more
realistic, to explore the usefulness of the mid-UV for estimating the
ages and metal abundances of old populations.

We discussed the spectral decomposition of an old population, which
shows that a mid-UV to optical/IR wavelength baseline allows good
separation of the main sequence turnoff from giant branch light.  The
turnoff and giant branch each contribute about 50\% of the V-band
light, but the turnoff dominates between 2500 and 4000 \AA.  Unlike the
optical region, spectral features in the mid-UV are not sensitive
to the dwarf/giant mixture in the population.  

Giant branch mass loss at the helium flash, which determines the
behavior of post-flash stars, is important in two ways:  high mass
loss produces hot phases (like the EHB stars), which strongly
influence the UV light; and very large amounts of mass loss
will reduce the cool RHB$+$AGB contribution to the infrared light.

We illustrate the general sensitivity of the mid-UV to age and metal
abundance using differential spectral diagrams.  There is a six
magnitude difference in the mid-UV continuum level (normalized at V)
over the metallicity range $-1.5 < \log Z/Z_{\odot} < +0.5$ and a
comparable difference (per unit $\log t$) for ages in the range 4-16
Gyr.  The rapid decline in flux below 3000 \AA\ in metal rich
populations may contribute to selection effects in the samples of
objects detected at high redshifts.  Sensitivity to age and abundance
declines longward of 4500 \AA, meaning that most of the spectral
information on old stellar populations resides below 4000 \AA. 

We derived a set of broad-band colors from our model spectra for the
range 2200 to 5500 \AA.  Logarithmic derivatives of mid-UV colors with
respect to age or metal abundance are 3--10 times larger than for the
UBV region.  Measurement of a single mid-UV color, such as $(25 - V)$, is
capable of placing a strong lower bound on the mean metallicity of an 
old population.  We investigated the ability of UBV and mid-UV
broad-band colors to separately determine age and abundance taking
into account observational precision in the color measurements.
Figure 7 shows that the mid-UV improves resolution in $\log t$, $\log
Z$ space by about a factor of 3 for a given observational precision.  
The ``age-metallicity degeneracy'' is significantly reduced for both
the continuum and spectral features in the mid-UV, but it cannot
be ignored even there. 

Contamination by the hot UVX (or ``UV-upturn") component is important
in the mid-UV. We confirm earlier investigations which found that this
can be very large, reaching over 80\% of the mid-UV light in galaxies
with the brightest far-UV sources.  In general it is straightforward
to remove its effects.  The true energy distribution of the cool
components, on which age and abundance estimates must be based, can be
very different from the uncorrected observations.  

We find that extinction should have relatively small effects on
parameters derived for old populations from mid-UV observations. 

Finally, we compared a 4 Gyr, solar abundance model derived from IUE
observations of nearby stars to the integrated mid-UV spectrum of the
Local Group elliptical galaxy M32, which has near-solar abundances.
The continuum and feature fit is excellent.  This shows that empirical
mid-UV stellar spectra can provide good fits to galaxy spectra (where
metallicities match) and indicates that the poorer results obtained from
theoretical spectra do arise from limitations of the synthesis models
for individual stars.  The importance of a concerted effort to improve
the fidelity of theoretical models for the emergent mid-UV spectra of
stars near the main sequence turnoff in old populations is clear.

\acknowledgments

We are grateful to Dave Burstein and Daniela Calzetti for providing
data, to Mike Fanelli for making the empirical mid-UV models, to Ruth
Peterson for many illuminating discussions, and to Jim Rose for helpful
suggestions.  This research was supported in part by NASA grant 5-6403.

\newpage

\singlespace


\begin{deluxetable}{cccccc}

\tablewidth{11.5cm}

\tablecaption{Synthetic Spectra Computed for this Study \label{ref-calcs}}
\tablehead{
\colhead{$Z$} &
\colhead{$\log Z/Z_{\odot}$} &
\colhead{[Fe/H]} & 
\colhead{[O/Fe]} &
\colhead{$Y_{\rm ZAMS}$}&
\colhead{Ages (Gyr)}}
\startdata

0.0006 & --1.5  & --1.48 & 0.60 & 0.235 & 2--20 \\
0.0060 & --0.5  & --0.47 & 0.23 & 0.250 & 2--20 \\
0.0169 &   0.0  &   0.00 & 0.00 & 0.270 & 2--20 \\
0.0400 &   0.3  &   0.38 & 0.00 & 0.270 & 2--20 \\
0.0600 &   0.5  &   0.58 & 0.00 & 0.270 & 2--20 \\
\enddata
\end{deluxetable}


\begin{deluxetable} {crrrrr}
\tablewidth{11cm}

\tablecaption{Isochrone Colors as a Function of Metallicity and Age \label{tab:color}}
\tablehead{
\colhead{t (Gyr)}
& \colhead{(22-V)}
& \colhead{(25-V)}
& \colhead{(30-V)}
& \colhead{(U-V)}
& \colhead{(B-V)}
}
\startdata
\cutinhead{(a): [Fe/H] =-1.48}
 2  &$  1.014 $&$  0.800 $&$  0.618 $&$  0.479 $&$  0.438 $\\
 3  &$  1.392 $&$  1.043 $&$  0.752 $&$  0.552 $&$  0.532 $\\
 4  &$  1.712 $&$  1.226 $&$  0.836 $&$  0.589 $&$  0.598 $\\
 5  &$  1.891 $&$  1.331 $&$  0.881 $&$  0.608 $&$  0.631 $\\
 6  &$  2.057 $&$  1.450 $&$  0.949 $&$  0.648 $&$  0.664 $\\
 7  &$  2.160 $&$  1.521 $&$  0.986 $&$  0.668 $&$  0.681 $\\
 8  &$  2.298 $&$  1.625 $&$  1.051 $&$  0.709 $&$  0.707 $\\
 9  &$  2.403 $&$  1.703 $&$  1.099 $&$  0.740 $&$  0.726 $\\
10  &$  2.453 $&$  1.735 $&$  1.114 $&$  0.748 $&$  0.732 $\\
11  &$  2.528 $&$  1.790 $&$  1.148 $&$  0.769 $&$  0.744 $\\
12  &$  2.603 $&$  1.846 $&$  1.183 $&$  0.792 $&$  0.755 $\\
13  &$  2.681 $&$  1.904 $&$  1.220 $&$  0.815 $&$  0.767 $\\
14  &$  2.763 $&$  1.966 $&$  1.260 $&$  0.840 $&$  0.780 $\\
15  &$  2.859 $&$  2.040 $&$  1.308 $&$  0.872 $&$  0.795 $\\
16  &$  2.932 $&$  2.095 $&$  1.345 $&$  0.897 $&$  0.807 $\\
17  &$  2.956 $&$  2.109 $&$  1.350 $&$  0.899 $&$  0.808 $\\
18  &$  2.983 $&$  2.126 $&$  1.359 $&$  0.904 $&$  0.810 $\\
19  &$  3.023 $&$  2.155 $&$  1.376 $&$  0.914 $&$  0.815 $\\
20  &$  3.079 $&$  2.197 $&$  1.404 $&$  0.933 $&$  0.823 $\\
\\
\cutinhead{(b): [Fe/H] =-0.47}
 2  &$  2.128 $&$  1.652 $&$  1.067 $&$  0.742 $&$  0.646 $\\
 3  &$  2.468 $&$  1.850 $&$  1.135 $&$  0.762 $&$  0.681 $\\
 4  &$  2.819 $&$  2.099 $&$  1.279 $&$  0.859 $&$  0.738 $\\
 5  &$  3.028 $&$  2.233 $&$  1.347 $&$  0.899 $&$  0.761 $\\
 6  &$  3.212 $&$  2.356 $&$  1.418 $&$  0.947 $&$  0.785 $\\
 7  &$  3.385 $&$  2.469 $&$  1.483 $&$  0.991 $&$  0.805 $\\
 8  &$  3.548 $&$  2.581 $&$  1.553 $&$  1.041 $&$  0.827 $\\
 9  &$  3.663 $&$  2.653 $&$  1.595 $&$  1.070 $&$  0.838 $\\
10  &$  3.806 $&$  2.747 $&$  1.653 $&$  1.111 $&$  0.855 $\\
11  &$  3.937 $&$  2.837 $&$  1.711 $&$  1.154 $&$  0.873 $\\
12  &$  4.051 $&$  2.913 $&$  1.760 $&$  1.190 $&$  0.887 $\\
13  &$  4.135 $&$  2.965 $&$  1.791 $&$  1.212 $&$  0.895 $\\
14  &$  4.208 $&$  3.006 $&$  1.813 $&$  1.226 $&$  0.900 $\\
15  &$  4.304 $&$  3.068 $&$  1.851 $&$  1.254 $&$  0.910 $\\
16  &$  4.398 $&$  3.129 $&$  1.889 $&$  1.281 $&$  0.920 $\\
17  &$  4.498 $&$  3.192 $&$  1.928 $&$  1.310 $&$  0.930 $\\
18  &$  4.604 $&$  3.258 $&$  1.968 $&$  1.339 $&$  0.940 $\\
19  &$  4.706 $&$  3.325 $&$  2.010 $&$  1.369 $&$  0.951 $\\
20  &$  4.824 $&$  3.400 $&$  2.057 $&$  1.404 $&$  0.963 $\\
\\
\cutinhead{(c): [Fe/H] = 0.00}
 2  &$  3.128 $&$  2.308 $&$  1.352 $&$  0.896 $&$  0.719 $\\
 3  &$  3.650 $&$  2.598 $&$  1.499 $&$  0.994 $&$  0.768 $\\
 4  &$  4.032 $&$  2.822 $&$  1.634 $&$  1.095 $&$  0.814 $\\
 5  &$  4.191 $&$  2.895 $&$  1.666 $&$  1.115 $&$  0.820 $\\
 6  &$  4.534 $&$  3.096 $&$  1.791 $&$  1.210 $&$  0.858 $\\
 7  &$  4.884 $&$  3.328 $&$  1.949 $&$  1.338 $&$  0.909 $\\
 8  &$  5.056 $&$  3.411 $&$  1.995 $&$  1.372 $&$  0.921 $\\
 9  &$  5.214 $&$  3.488 $&$  2.034 $&$  1.398 $&$  0.928 $\\
10  &$  5.383 $&$  3.582 $&$  2.091 $&$  1.441 $&$  0.942 $\\
11  &$  5.537 $&$  3.669 $&$  2.143 $&$  1.479 $&$  0.955 $\\
12  &$  5.698 $&$  3.760 $&$  2.197 $&$  1.520 $&$  0.968 $\\
13  &$  5.859 $&$  3.855 $&$  2.254 $&$  1.563 $&$  0.982 $\\
14  &$  6.010 $&$  3.946 $&$  2.309 $&$  1.604 $&$  0.995 $\\
15  &$  6.152 $&$  4.035 $&$  2.363 $&$  1.644 $&$  1.008 $\\
16  &$  6.300 $&$  4.131 $&$  2.421 $&$  1.687 $&$  1.022 $\\
17  &$  6.458 $&$  4.237 $&$  2.485 $&$  1.734 $&$  1.038 $\\
18  &$  6.604 $&$  4.338 $&$  2.547 $&$  1.780 $&$  1.053 $\\
19  &$  6.698 $&$  4.396 $&$  2.577 $&$  1.800 $&$  1.058 $\\
20  &$  6.793 $&$  4.455 $&$  2.608 $&$  1.821 $&$  1.064 $\\
\\
\cutinhead{(d): [Fe/H] = 0.38}
 2  &$  4.499 $&$  3.008 $&$  1.665 $&$  1.094 $&$  0.783 $\\
 3  &$  5.231 $&$  3.414 $&$  1.918 $&$  1.296 $&$  0.865 $\\
 4  &$  5.706 $&$  3.692 $&$  2.096 $&$  1.439 $&$  0.919 $\\
 5  &$  6.086 $&$  3.904 $&$  2.219 $&$  1.532 $&$  0.949 $\\
 6  &$  6.214 $&$  3.986 $&$  2.267 $&$  1.568 $&$  0.960 $\\
 7  &$  6.359 $&$  4.068 $&$  2.310 $&$  1.597 $&$  0.967 $\\
 8  &$  6.555 $&$  4.195 $&$  2.384 $&$  1.653 $&$  0.984 $\\
 9  &$  6.741 $&$  4.328 $&$  2.467 $&$  1.717 $&$  1.006 $\\
10  &$  6.898 $&$  4.430 $&$  2.518 $&$  1.751 $&$  1.014 $\\
11  &$  7.135 $&$  4.622 $&$  2.639 $&$  1.845 $&$  1.044 $\\
12  &$  7.308 $&$  4.767 $&$  2.733 $&$  1.919 $&$  1.071 $\\
13  &$  7.368 $&$  4.798 $&$  2.736 $&$  1.916 $&$  1.066 $\\
14  &$  7.464 $&$  4.867 $&$  2.769 $&$  1.936 $&$  1.070 $\\
15  &$  7.591 $&$  4.969 $&$  2.825 $&$  1.977 $&$  1.082 $\\
16  &$  7.739 $&$  5.095 $&$  2.899 $&$  2.033 $&$  1.099 $\\
17  &$  7.870 $&$  5.206 $&$  2.961 $&$  2.078 $&$  1.113 $\\
18  &$  7.982 $&$  5.299 $&$  3.009 $&$  2.112 $&$  1.123 $\\
19  &$  8.078 $&$  5.380 $&$  3.048 $&$  2.138 $&$  1.129 $\\
20  &$  8.182 $&$  5.469 $&$  3.094 $&$  2.170 $&$  1.138 $\\
\\
\cutinhead{(e): [Fe/H] = 0.58}
 2  &$  5.349 $&$  3.442 $&$  1.898 $&$  1.277 $&$  0.839 $\\
 3  &$  6.033 $&$  3.848 $&$  2.152 $&$  1.481 $&$  0.915 $\\
 4  &$  6.561 $&$  4.219 $&$  2.399 $&$  1.682 $&$  0.989 $\\
 5  &$  6.766 $&$  4.344 $&$  2.456 $&$  1.718 $&$  0.994 $\\
 6  &$  6.944 $&$  4.468 $&$  2.524 $&$  1.766 $&$  1.007 $\\
 7  &$  7.121 $&$  4.599 $&$  2.599 $&$  1.821 $&$  1.023 $\\
 8  &$  7.303 $&$  4.741 $&$  2.681 $&$  1.883 $&$  1.042 $\\
 9  &$  7.520 $&$  4.915 $&$  2.783 $&$  1.959 $&$  1.066 $\\
10  &$  7.688 $&$  5.050 $&$  2.857 $&$  2.013 $&$  1.082 $\\
11  &$  7.828 $&$  5.161 $&$  2.913 $&$  2.052 $&$  1.091 $\\
12  &$  7.972 $&$  5.280 $&$  2.978 $&$  2.097 $&$  1.104 $\\
13  &$  8.113 $&$  5.399 $&$  3.041 $&$  2.142 $&$  1.117 $\\
14  &$  8.262 $&$  5.528 $&$  3.108 $&$  2.189 $&$  1.130 $\\
15  &$  8.399 $&$  5.650 $&$  3.173 $&$  2.236 $&$  1.144 $\\
16  &$  8.493 $&$  5.731 $&$  3.209 $&$  2.259 $&$  1.149 $\\
17  &$  8.585 $&$  5.810 $&$  3.243 $&$  2.280 $&$  1.153 $\\
18  &$  8.696 $&$  5.913 $&$  3.294 $&$  2.316 $&$  1.163 $\\
19  &$  8.810 $&$  6.025 $&$  3.355 $&$  2.361 $&$  1.176 $\\
20  &$  8.912 $&$  6.126 $&$  3.408 $&$  2.399 $&$  1.187 $\\
\enddata
\end{deluxetable}


\begin{deluxetable}{cccc}

\tablewidth{10.5cm}

\tablecaption{Photometric Derivatives at  $t = 10\,$ Gyr, $Z = Z_{\odot}$ 
\label{ref-photder}}

\tablehead{
\colhead{Color} &
\colhead{$\partial\, C / \partial \log t$} &
\colhead{$\partial\, C / \partial \log Z$} &
\colhead{$\partial\, C / \partial E(B-V)$} }

\startdata

$22 - V$  & 3.57 & 3.36 & 6.11\\
$25 - V$  & 2.06 & 1.78 & 4.31\\
$30 - V$  & 1.23 & 0.93 & 3.07\\
$U - V$   & 0.93 & 0.70 & 1.73\\
$B-  V$   & 0.34 & 0.19 & 1.00\\

\enddata
\end{deluxetable}


\begin{deluxetable}{cccccc}

\tablewidth{32pc}

\tablecaption{UVX Contamination in Selected Galaxies}

\tablehead{
\colhead{Object} &
\colhead{$(15-V)_0$} &
\colhead{$(25-V)_0$} &
\colhead{$p(V)$} &
\colhead{$p^{\prime}(25)$} &
\colhead{$(25-V)_{\rm COOL}$}  } 

\startdata

NGC221/M32 & 4.50 & 3.25 & 0.05\% & 9.8\% & 3.37 \\
NGC224/M31 & 3.46 & 3.73 & 0.12\% & 39\%  & 4.30 \\
NGC4552/M89 & 2.32 & 3.12 & 0.36\% & 65\% & 4.25 \\
NGC4649/M60 & 2.20 & 3.27 & 0.40\% & 82\% & 5.16 \\

\enddata
\end{deluxetable}


\begin{deluxetable}{cc}

\tablewidth{7cm}

\tablecaption{Components of M32 Empirical Model}

\tablehead{
\colhead{Stellar Group\tablenotemark{a}} &
\colhead{Percentage V-Light} }

\startdata

 F5-7 V  &   10.8\% \\ 
 F8-9 V  &    7.4\% \\ 
 G0-5 V  &    8.2\% \\ 
 G6-9 V  &    2.9\% \\ 
 K0-1 V  &    3.1\% \\ 
 K2-3 V  &    1.5\% \\ 
 K5-M0 V &    3.2\% \\ 
 F2-7 IV &    3.1\% \\ 
 G0-2 IV &    3.9\% \\ 
 G5-8 IV &    4.1\% \\ 
 G8-K1 IV &   9.0\% \\ 
 G5-K0 III &  6.0\% \\ 
 K0-1 III &  12.4\% \\ 
 K2 III   &  11.2\% \\ 
 K3 III   &   3.4\% \\ 
 K4-5 III &   2.6\% \\ 
 K7-M3 III &  7.2\% \\ 
       
\enddata

\tablenotetext{a}{Group mean spectra from Fanelli et al.\ (1992)}

\end{deluxetable}


\epsscale{0.8}

\begin{figure}[t]
\plotone{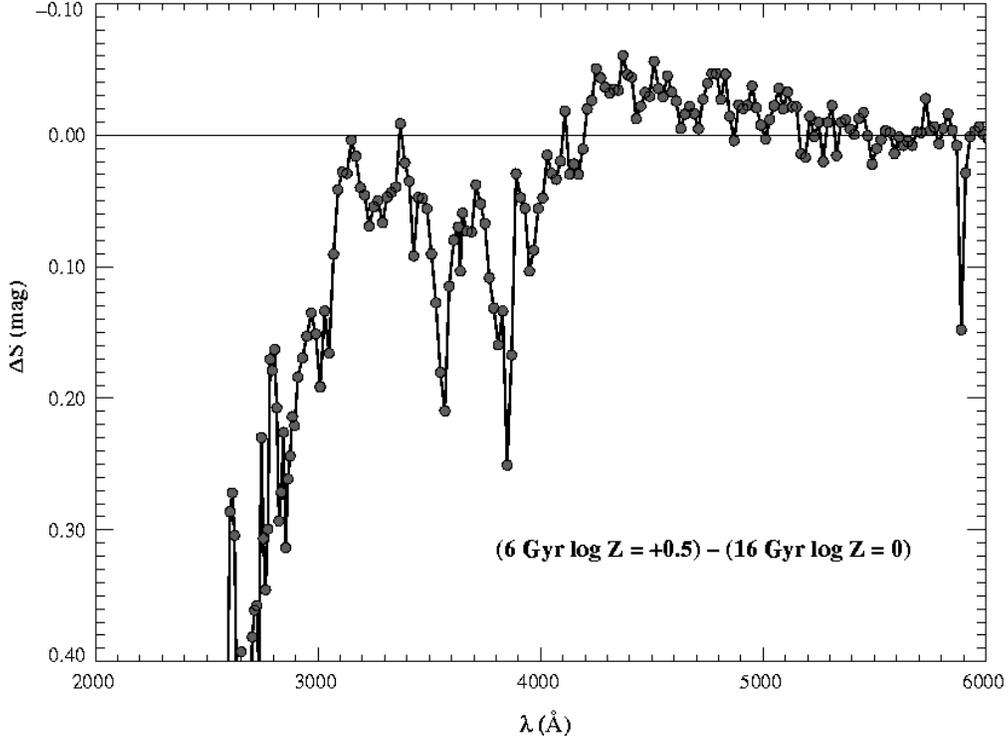}
\vspace{10pt}
\caption{{\small\protect\label{pair-diff} {\singlespace
The difference between the spectra of a model at 16 Gyr, $\log
Z/Z_{\odot} = 0,$ and a model at 6 Gyr, $\log Z/Z_{\odot} = +0.5.$ The
original spectra were normalized at the $V$-band.  This case of
age/metallicity degeneracy is cited by Worthey (1994) as difficult to
distinguish on the basis of optical colors.  The models in this plot
confirm that conclusion for the spectral range 4500--6000 \AA\
considered by Worthey (apart from the Na I doublet at 5890 \AA).  But
large differences between the models appear in both the continuum and
line strengths at wavelengths below 4200 \AA\ and increase toward the
mid-UV.  }}}
\end{figure}

\begin{figure}[ht]
\plotone{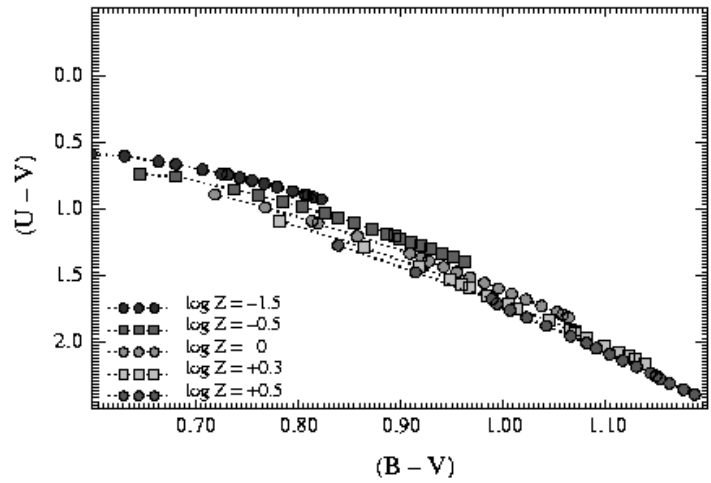}
\vspace{10pt}
\caption{\protect\label{uvbv}
The model two-color optical $(U-V),$ $(B-V)$ diagram as a function of age
and metallicity.  Model isochrones have been combined with Kurucz
(1992) atmospheres as described in the text.  
The range of ages plotted is 2--20 Gyr with the 20 Gyr
models at the right-hand (red) end of the sequence; the models are
1 Gyr apart.   Metallicity labels give values with respect to solar
abundance.  }
\end{figure}

\begin{figure}[ht]
\plotone{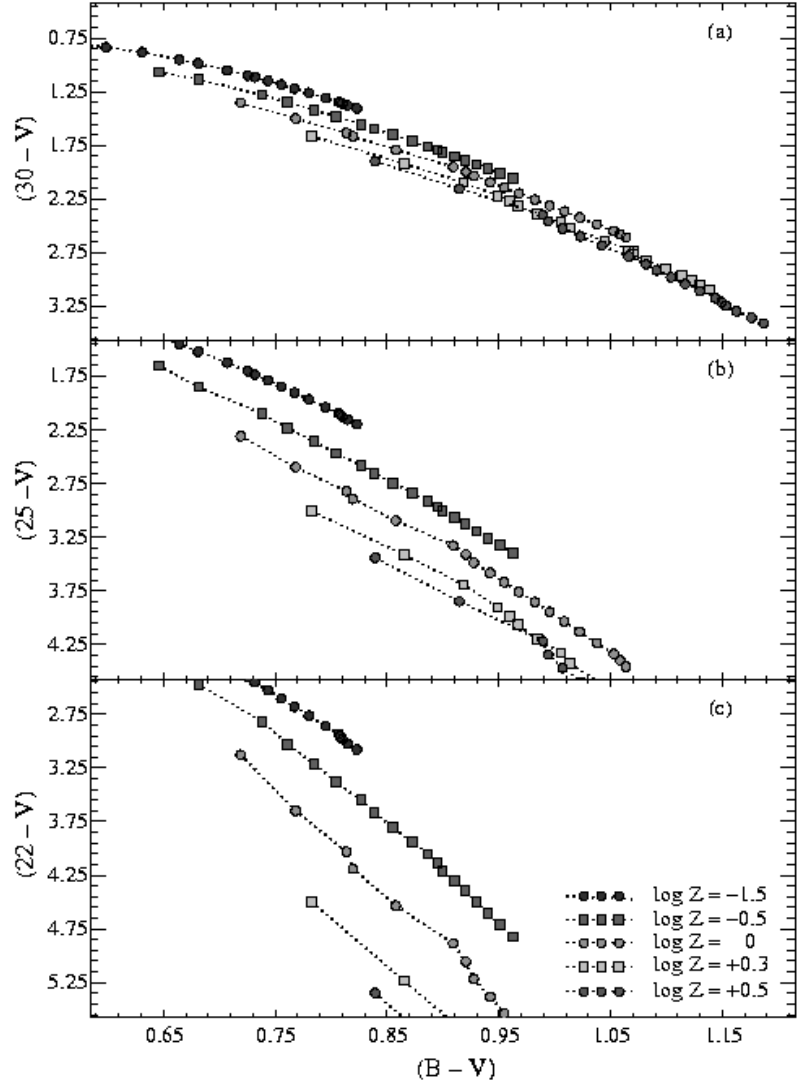}
\vspace{10pt}
\caption{\protect\label{amd} 
Model two-color diagrams generated for the HST/WFPC2 mid-UV colors.
The plots are drawn to the same scale as Fig.~\protect\ref{uvbv},
showing the improved separation of models of different metallicity at
UV wavelengths. Metallicity labels give values with respect to solar
abundance. }
\end{figure}

\begin{figure}[ht]
\plotone{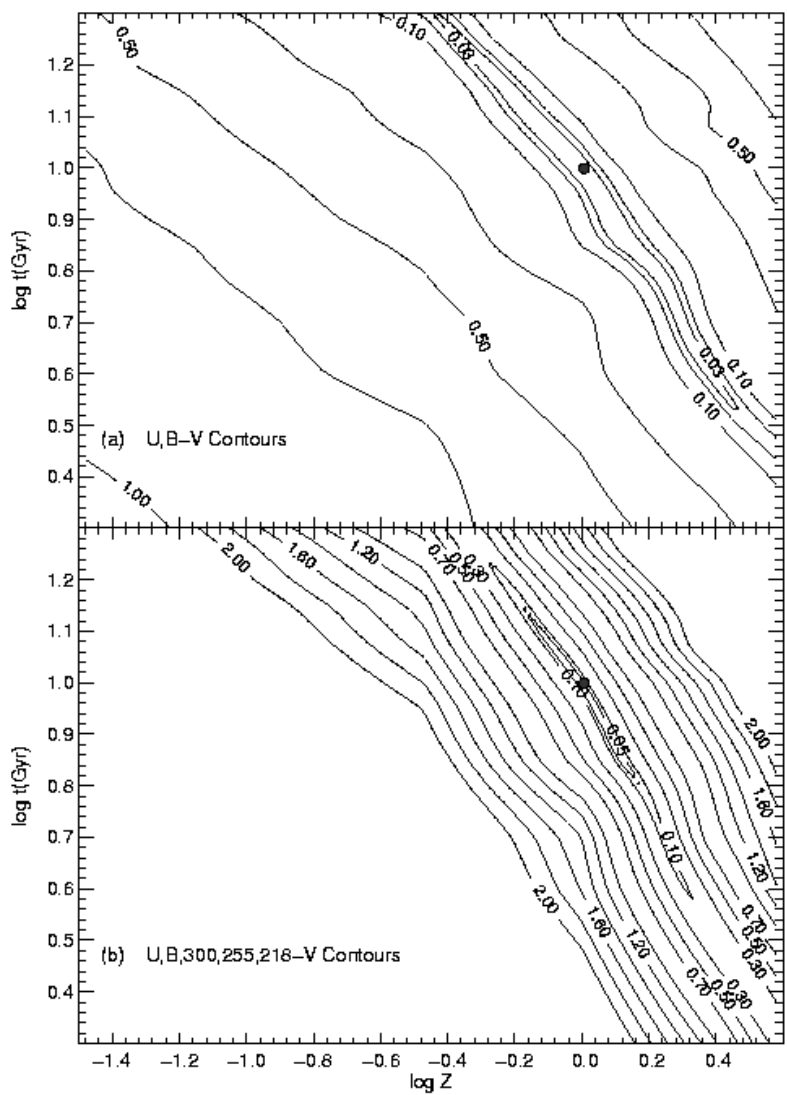}
\vspace{10pt}
\caption{ \protect\label{ellipse}
The sensitivity of inferred ages and metallicities to observational
error. Plotted in the age/metallicity plane are contours of
root-mean-square color differences between the theoretical model at
10 Gyr, $Z_{\odot}$ (filled circle) and surrounding models. The metallicity
axis is normalized at solar abundance.  For the
optical two-color plot (panel a), the innermost contour represents an
rms difference of 0.03 mag, and spans a large age/metallicity range.
Panel (b) shows the same plane, but with 3 mid-UV broadband colors
added.}
\end{figure}

\begin{figure}[ht]
\plotone{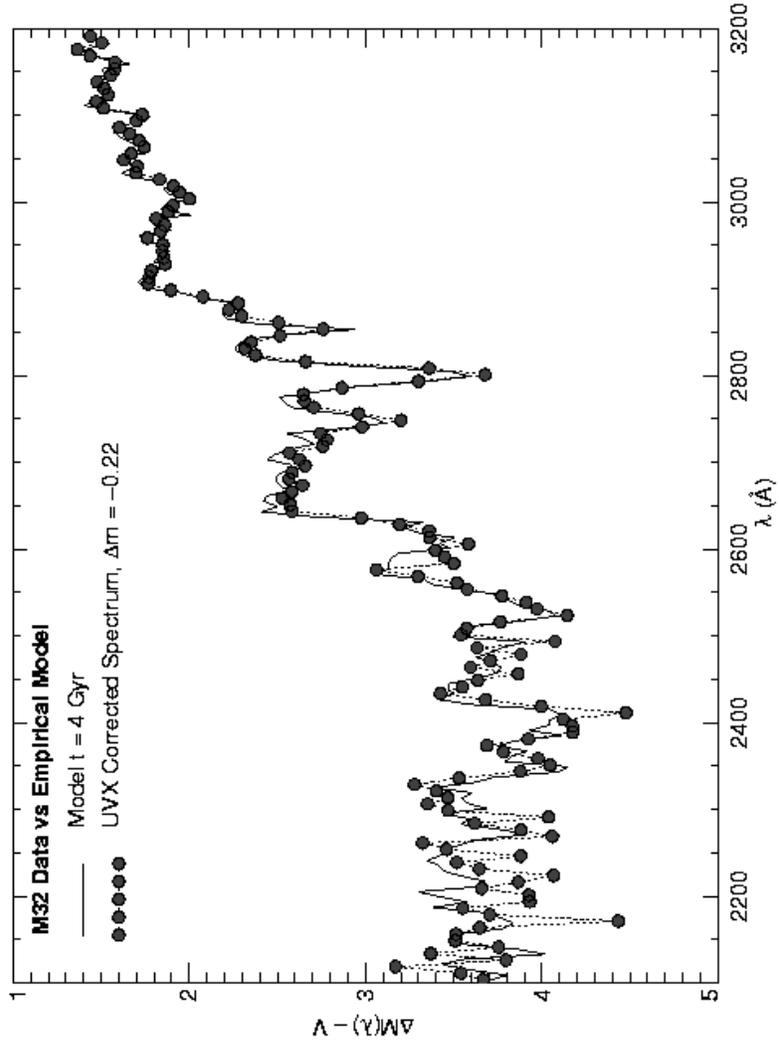}
\vspace{10pt}
\caption{\protect\label{m32} 
Comparison between the IUE mid-UV spectrum of the center of the Local Group
elliptical M32, corrected for UVX contamination, and a 4 Gyr old,
solar abundance model based on observed spectra of nearby bright
stars.  See text for details.
}
\end{figure}

\end{document}